# All-Polarization Maintaining Single-Longitudinal-Mode Fiber Laser with Ultra-High OSNR, Sub-kHz Linewidth and Extremely High Stability

ZHENGKANG WANG,[1] JIANMING SHANG,[2] SIQIAO LI,[1] KUANLIN MU,[1] SONG YU,[2] AND YAOJUN QIAO,[1,*]

[1]*Beijing Key Laboratory of Space-Ground Interconnection and Convergence, School of Information and Communication Engineering, Beijing University of Posts and Telecommunications, 10 Xitucheng Road, Beijing, 100876, China*
[2]*School of Electronic Engineering, Beijing University of Posts and Telecommunications, 10 Xitucheng Road, Beijing, 100876, China*
*\*qiao@bupt.edu.cn*

**Abstract:** An all-polarization maintaining (PM) single-longitudinal-mode (SLM) erbium-doped fiber laser (EDFL) with ultra-high optical signal-to-noise ratio (OSNR), ultra-narrow linewidth and extremely high stability is proposed and experimentally demonstrated. A doubling-ring passive subring resonator (DR-PSR) composed of two single-coupler fiber rings and a length of unpumped EDF-based saturable absorber filter is designed and employed in the EDFL to serve as the efficient SLM selecting element to guarantee SLM lasing with excellent output performance. The all-PM structure enables the proposed EDFL to present strong ability to resist the environment disturbance. At the pump power of 100 mW, we obtain an SLM EDFL with an ultra-high OSNR of 83 dB and an ultra-narrow linewidth of 459 Hz. For the SLM operation, the all-PM EDFL processes outstanding stability performance of both the wavelength lasing and the output power. The maximum fluctuations of the center wavelength and output power are 0.012 nm and 0.01 dB.

## 1. Introduction

Single-longitudinal-mode (SLM) fiber lasers possess the inherent merits of high beam quality, low noise, narrow width and excellent capability with fiber systems. SLM erbium-doped fiber lasers (EDFLs) with high optical signal to noise ratio (OSNR), narrow linewidth, high stability and easy operability have attracted a great deal of attention for their potential applications in optical communication, optical metrology, coherent beam combining technology and high-resolution spectroscopy [1-4]. At present, the cavities of EDFLs mainly include three schemes: the distributed feedback [5], the distributed Bragg reflection [6], and the ring cavity [7-8]. Among them, the ring cavity SLM EDFL is acclaimed for its ability to avoid the spatial-hole burning induced by the standing-wave effect presented in the linear cavity configuration [9]. Additionally, the ring cavity EDFL uses a relatively long gain medium and long cavity design, which is helpful to obtain a high output power and a narrow linewidth [10]. However, it is difficult to obtain SLM EDF ring laser with outstanding performance due to the unavoidable multi-longitudinal mode (MLM) oscillation or mode hopping, which is mainly caused by the homogeneous broadening of the EDF and the long cavity length of the EDFL [11]. To achieve stable SLM lasing for EDF ring lasers, researchers have utilized substantial techniques and optical components in the cavity, such as the multiple-ring compound resonator [7-8, 12-13], the unpumped EDF-based saturable absorber (SA) filter [14], the Mazh-zehnder interferometer [15], the ultra-narrow pass-band filter [16], the phase shifted fiber Bragg grating (FBG) [17], the micro-ring resonator filter [18] and the micro-sphere filter [19].

Using a multiple-ring compound resonator is an effective way to obtain the SLM operation of the ring cavity EDFL by producing an effective wide free spectrum range (FSR) based on the Vernier effect. Several multiple-ring compound resonators have been employed in the EDFLs for the stable SLM output. Nevertheless, there are certain weaknesses with some multiple-ring constructions, including the relatively large linewidths of the EDFLs [7, 12], the insufficient stability and the unsatisfactory OSNRs [9, 12], as well as the strictly required value of the length of the multiple-ring resonators [7,13]. In order to overcome the shortcomings of the above EDFLs, it is useful to employ the unpumped EDF-based SA filter in the EDF ring laser. An ultra-narrowband self-tracking filter can be established in the unpumped EDF to stabilize the SLM operation and narrow the linewidth of the EDFL under the specific standing-wave condition [14]. In our previous work, we have designed and manufactured an SLM EDFL with excellent stability performance by cooperating a multiple-ring compound resonator with an SA filter [8]. But its need of a polarization controller (to adjust the polarization state of the EDFL to obtain the optimal SLM operation) and its sensitivity to external perturbations made it difficult for our last EDFL to meet the requirement of some important applications. Moreover, the linewidth and OSNR performance of the proposed EDFL are also insufficient for important applications.

In fact, the mode-hopping of the EDFL can originate from the external environment disturbance to the relatively long cavity length of the ring cavity. To resist the environment disturbance and prolong the mode-hopping-free SLM operation time of the EDF ring laser, some techniques have been used in the EDFL, such as the vibration isolation and the thermal insulation. Yet, it remains difficult to meet the needs of practical applications. The polarization maintaining (PM) fibers can significantly improve the ability of the fiber laser to resist environmental disturbances, such as vibration, temperature changes and the move of fibers. If all the fibers used in the SLM EDFL are PM fibers, the all-PM design can substantially improve the stability of the SLM operation and increase the practicality of the EDFL. In fact, all-PM design has been extensively used in mode-locked fiber laser [20]. However, to the best of authors' knowledge, there has been almost no research on the study of all-PM SLM EDFL. Therefore, developing all-PM SLM EDFL with excellent output performance and with strong ability to resist the environment disturbance is of great significance.

In this paper, we propose and demonstrate an all-PM EDF ring laser based on a high-quality double-ring passive subring resonator (DR-PSR). The DR-PSR, composed of two single-coupler fiber rings and a length of an unpumped EDF, serves as the core SLM selecting element of the proposed EDFL. The two single-coupler fiber rings function as effective mode filters to suppress the dense MLM oscillation of the EDFL to benefit the SLM selection. The unpumped EDF, serving as the SA filter, is utilized to guarantee and stabilize the SLM operation of the EDFL. In addition, because of the all-PM structure, the proposed SLM EDFL presents excellent capability to resist the environment disturbance. With the excellent mode selecting capability of the DR-PSR and the all-PM design, the proposed EDFL demonstrates outstanding performance of ultra-high OSNR, ultra-narrow linewidth and extremely high stability.

## 2. Experimental setup and principles

### 2.1 Experimental configuration of the all-PM EDFL

Fig. 1 illustrates the schematic of the proposed all-PM EDF ring laser, which mainly consists of a segment of a 2-m-long PM EDF1 (IXF-EDF-FGL-PM-L2), a PM circulator (CIR1), a PM FBG and a DR-PSR. The PM EDF1 is used as the gain medium pumped by a 980 nm laser diode (LD) with 200 mW maximum output power through a PM 980/1550 nm wavelength division multiplexer (WDM). The 1550.12 nm PM FBG with >90% reflectivity and 0.1 nm 3-dB bandwidth functions as a reflection filter to complete the original wavelength selection and to reduce the potential longitudinal mode in the main cavity by cooperating with the PM CIR1. The CIR1 guarantees the unidirectional propagation of the light wave in the main cavity to avoid the MLM oscillation caused by the spatial-hole burning. The generated laser is extracted by the third port of PM optical coupler (OC1) with a 50:50 coupling ratio, and a PM isolator (ISO) spliced to the output port is used to suppress the external reflection.

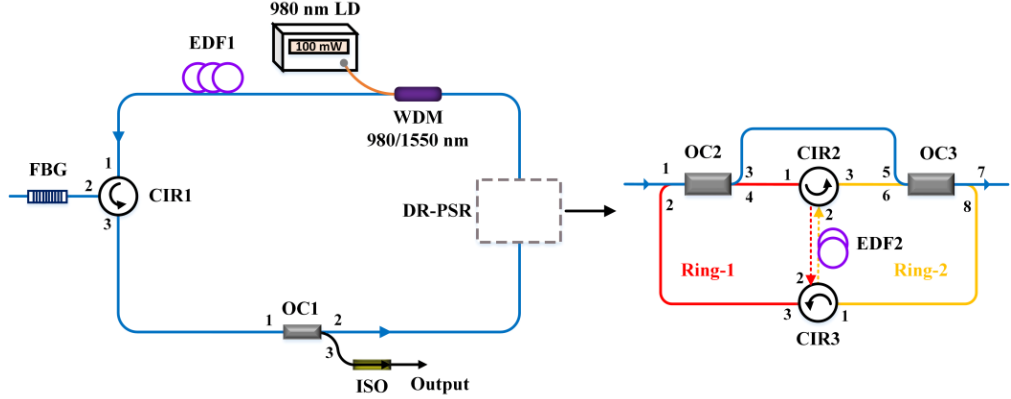

Fig. 1 Experimental setup of the designed all-PM EDFL. OC, optical coupler; LD, laser diode; WDM, wavelength division multiplexer; EDF, erbium-doped fiber; FBG, fiber Bragg grating; CIR, circulator; ISO, isolator; DR-PSR, double-ring passive subring resonator.

The DR-PSR inserted between the pass port of the PM WDM and the second port of the PM OC1 is constructed with a pair of PM optical couplers (OC2 and OC3) with coupling ratios of 90:10, a pair of PM circulators (CIR2 and CIR3) and a segment of 2.5-m-long unpumped EDF2 (IXF-EDF-FGL-1480-PM). When the input light beams are injected into OC2, 10% of the output light beams go straight into the fifth port of the OC3, and 90% of the output light beams go into the first port of the CIR2. For the light beams going into the first port of the CIR2, the output light beams from the second port of CIR2 go into the second port of CIR3 through the EDF2, and the output light beams from the third port of CIR3 feed back into the second port of OC2. As shown in Fig. 1, OC2, EDF2 and two CIRs form a single-coupler fiber ring as marked in Ring-1 (red line). As for the light beams going into OC3, 10% of the output light beams served as the output of DR-PSR while 90% of the output light beams travel into the fifth port of CIR3. The output light beams from CIR3 pass through the EDF2 and go into the second port of CIR2. Then, the output light beams from the third port of CIR2 go into the sixth port of OC3. Here, OC3, EDF2 and two CIRs form another single-coupler fiber ring as marked in Ring-2 (yellow line).

*2.2 SLM selecting mechanism based on the DR-PSR*

The length of the main cavity is about 13.9 m corresponding to a longitudinal-mode spacing of 14.7 MHz, and the 3-dB bandwidth of the PM FBG is ~0.1 nm corresponding to a frequency bandwidth of 12.5 GHz at 1550 nm. In order to obtain the SLM operation, the proposed DR-PSR must guarantee that only one longitudinal mode to be selected from more than 800 modes. As shown in Fig. 2, the DR-PSR consists of two single-longitudinal fiber rings (Ring-1 and Ring-2) with each of the length of 3.7 m and 4.3 m, and the corresponding FSRs of Ring-1 and Ring-2 can be calculated as 55.1 MHz and 47.4 MHz according to the equation:

$$FSR = c/nL \qquad (1)$$

Where $c$ is the light speed in vacuum, $n \approx 1.47$ is the effective refractive index at 1550 nm of single-mode fiber and $L$ is the length of the single-longitudinal fiber ring. Then the effective FSR of the DR-PSR can be calculated to be ~2.6 GHz in accordance with the Vernier effect [21].

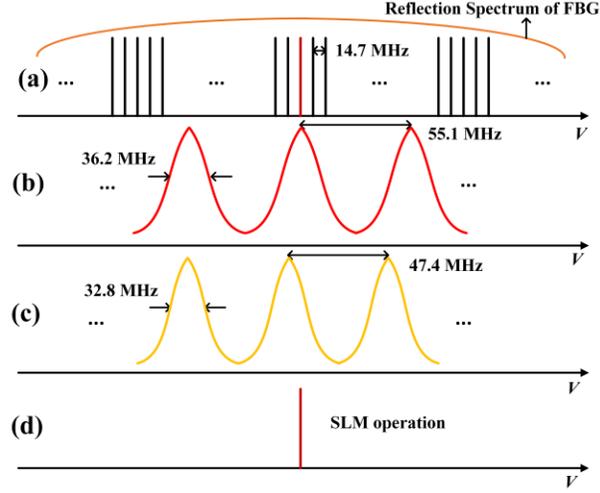

Fig. 2 Schematic diagram of the SLM operation in the EDFL. (a) Dense longitudinal modes of the main ring cavity and the reflection spectrum of the FBG; (b) The FSR and the 3-dB pass-band of the Ring-1; (c) The FSR and the 3-dB pass-band bandwidth of the Ring-2; (d) The selected SLM.

The two single-coupler fiber rings formed in the DR-PSR can decrease the number of longitudinal modes and suppress the dense MLM oscillation in the EDFL effectively. However, the effective FSR of the DR-PSR is much smaller than the 3-dB reflecting bandwidth of the PM FBG; there are more than one pass band located in the 3-dB bandwidth of the FBG. Moreover, as shown in Fig. 2(b) and Fig. 2(c), the 3-dB bandwidth of Ring-1 resonator and Ring-2 resonator are 36.2 MHz and 32.8 MHz respectively. As the 3-dB bandwidth of Ring-1 resonator or Ring-2 resonator is bigger than the longitudinal-mode spacing of the main cavity, there are more than one longitudinal mode that can pass through the DR-PSR.

The bandwidth of the single-coupler fiber ring (take Ring-1 for example) can be calculated according to the transfer function of Ring-1:

$$T = \left|\frac{E_3}{E_1}\right|^2 = \frac{A^2[(1-k) - 2A\sqrt{1-k}\cos(2\pi nvl/c)e^{-\alpha l} + A^2 e^{-2\alpha l}]}{1 - 2A\sqrt{1-k}\cos(2\pi nvl/c)e^{-\alpha l} + A^2(1-k)e^{-2\alpha l}} \quad (2)$$

Where $E_3$ is the output field of Ring-1, $E_1$ is the input field of Ring-1, $A = \sqrt{1-\gamma}$ and $\gamma$ is the coupling loss of the OC, $k = 0.9$ is the splitting ratio of the OC and $n \approx 1.47$ is the effective group refractive index at 1550 nm. In addition, $\alpha$ is the transmission loss of the fiber and $l = 3.7\ m$ is the length of Ring-1.

As mentioned in the "Experimental configuration of the all-PM EDFL", two counter-propagating light beams travel through the EDF2 in the DR-PSR. Here, the unpumped EDF2 serves as the SA filter to guarantee and stabilize the SLM operation. A self-tracking filter with ultra-narrow bandwidth can form in the unpumped EDF2 according to the theories of standing wave and nonlinear optics, and the full width at half maximum (FWHM) of the self-tracking filter is calculated to be ~7.5 MHz according to the following equation [22]:

$$\Delta f = \frac{c}{\lambda} \frac{2\Delta n}{n_{eff} \lambda} \sqrt{(\frac{\Delta n}{2n_{eff}})^2 + (\frac{\lambda}{2n_{eff} L_g})^2} \quad (3)$$

where $\Delta n$ is the variation of the refractive index, which can be given by the Kramers-Kronig relation [23]. $n_{eff}$ is the effective index of the unpumped EDF2, $\lambda$ is the central wavelength and $L_g$ is the length of the unpumped EDF2 in the DR-PSR.

Because the 7.5 MHz FWHM of the self-tracking filter is smaller than the longitudinal-mode spacing of the main cavity, the SLM operation of the EDFL can be guaranteed eventually. According to the analysis above, the proposed DR-PSR serves as the core SLM selecting element to guarantee the SLM output of the EDFL at any time, as shown in Fig. 2(d).

### 3. Experimental results and discussions

In this paper, we carried out all experiments at the laboratory temperature without temperature compensation or vibration isolation used for the SLM EDFL. Since the EDFL output only one polarization state, which benefited from the all-PM structure, it was unnecessary to adjust the polarization state by the polarization controller. The proposed EDFL presented an ultra-high OSNR, an ultra-narrow linewidth and an outstanding stability performance, which are shown as follows.

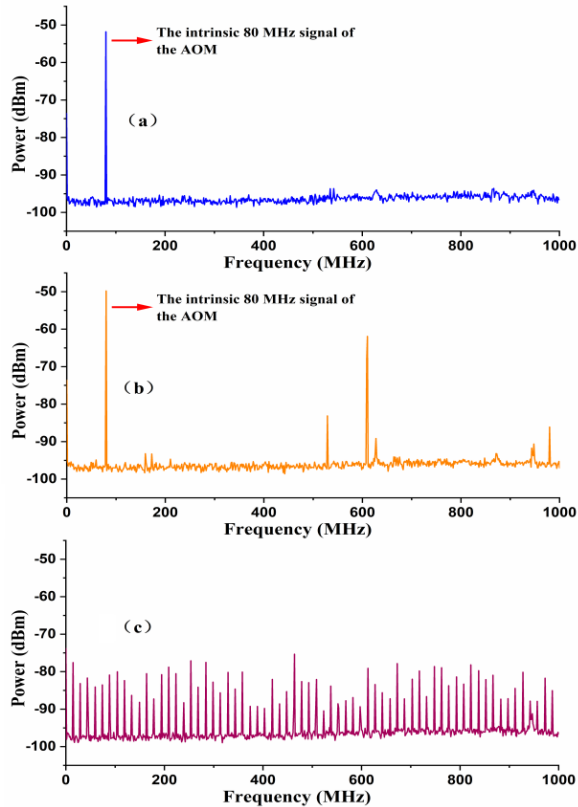

Fig. 3. The radio frequency beating spectra measured by delayed self-heterodyne measurement system in: (a) 0~1 GHz span; (b) 0~1 GHz span with unpumped EDF replaced by an identical long length SMF; (c) 0~1 GHz span with DR-PSR replaced by an identical long length SMF.

Firstly, the SLM output performance of the EDFL and the SLM selecting capability of the DR-PSR were verified through the self-heterodyne system when the pump power was fixed at 100 mW. The self-heterodyne system was composed of a 1 GHz photodetector (PD), a RF electrical spectrum analyzer (Rohde & Schwarz FSV), a Mach-Zehnder interferometer with an 80 MHz acoustic optical modulator (AOM) and a 47 km single mode fiber delay line in two arms respectively. Fig. 3(a) shows the RF spectrum measured in 0~1 GHz range with 100 kHz resolution bandwidth (RBW). The results show that there are no other obvious beating signals captured except for the intrinsic 80 MHz sifted signal of the AOM. Moreover, there was no apparent mode hopping captured of the all-PM EDFL for 2 hours continuous monitoring, indicating that the proposed EDFL was in stable SLM operation. In order to investigate the SLM selecting capability of the SA filter formed in the DR-PSR, we replaced the 2m-long EDF2 with a 2m-long SMF, and there were some beating signals captured in the RF spectrum of the EDFL (as shown in Fig. 3(b)). Then, we replaced the entire DR-PSR with a length of SMF to maintain the original length of the main cavity, and it turned out that there were many peaks in the RF spectrum of the because of the dense MLM operation, as shown in Fig. 3(c). Fig. 3(b) and Fig. 3(c) verify the excellent SLM selecting capability of the DR-PSR, and the experimental results are consistent with the principle analysis in "Experimental Principles". In addition, as it can be seen in Fig. 3(c), the frequency spacing of the adjacent peaks is ~15 MHz, which is consistent with the value of the longitudinal-mode spacing of the main cavity of the proposed EDFL.

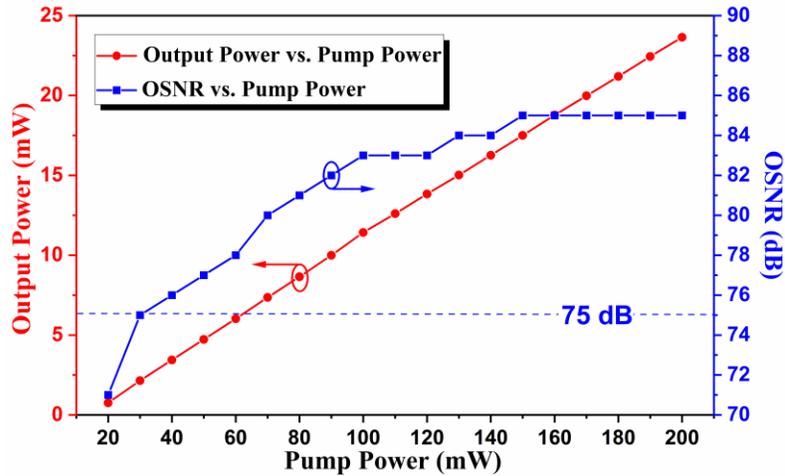

Fig. 4 Output power and OSNR variations versus pump power

Then, the relationships between the output power and OSNR variations versus the pump power were measured by a power meter (JW3223C) with the resolution of 0.01 dB and an optical spectrum analyzer (OSA, AQ637OD) with the resolution of 0.02 nm, as shown in Fig. 4. As Fig. 4 shows, the threshold pump power of the proposed EDFL was about 20 mW with the slope efficiency of ~13.1%. In addition, the output power of the EDFL reached 23.63 mW with no observed output saturation at the pump power of 200 mW, indicating that the output power of the SLM EDFL could be further increased if the pump power had been big enough. As for the OSNR value of the EDFL, the proposed EDFL could output SLM laser with OSNR of >75 dB when the pump power was bigger than 30 mW. With the increase of the pump power from 30 mW to 200 mW, the OSNRs of the EDFL had an increasing trend up to ~85 dB, indicating the excellent mode selecting performance of the DR-PSR. However, the OSNR value of the EDFL tends to saturate when the pump power reached 100 mW. In this paper, the optical spectrum, the stability, the linewidth and the polarization performance of the proposed SLM EDFL were further investigated at the pump power of 100 mW, which are demonstrated as follows.

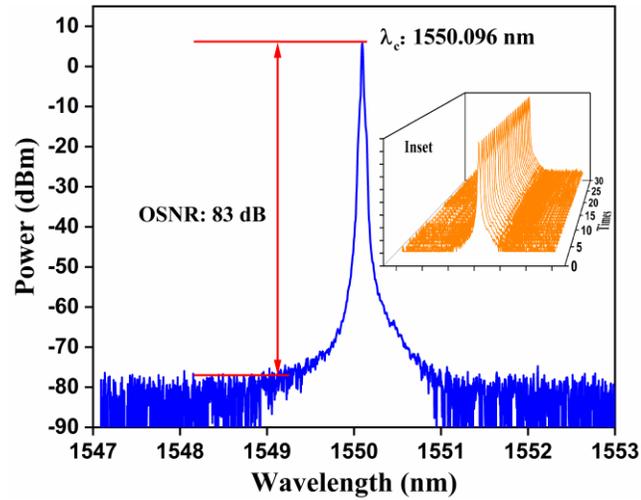

Fig. 5 Optical spectrum measured by the OSA at a pump power of 100 mW. Inset displaying 30 repeated OSA scans at 1-min intervals.

The optical spectrum of the EDFL was measured by the OSA, which is shown in Fig. 5. The center wavelength of the EDFL was 1550.096 nm with an ultra-high OSNR of 83 dB. As can be seen, the lasing wavelength of the EDFL was slightly different from the center reflecting wavelength of the PM FBG, which was mainly caused by the thermal effect and vibration at the PM FBG. Additionally, the spectra of 30 times repeated OSA scans at 1 min interval are shown in the inset of Fig. 3, exhibiting stable operation without significant variations in either the output power or lasing wavelength. (可以分析一下为啥 OSNR 值很高)

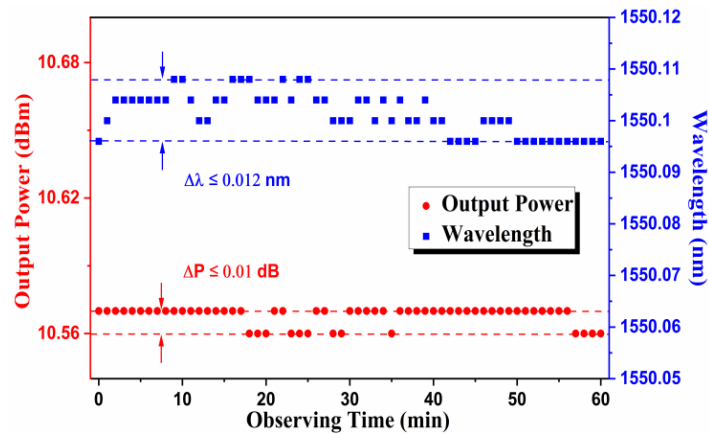

Fig. 6 The stability performances of the output power and the lasing wavelength at pump power of 100 mW in a 1-min interval over 1 h

To further verify the extremely high stability of the proposed EDFL, we investigated the variations of the output power and lasing wavelength through an OSA and a power meter. The stability of the EDFL was investigated during a 1-hour experiment with the interval of 1-min, as shown in Fig. 6. As can be seen from Fig. 6, the maximum wavelength variation of the EDFL was 0.012 nm, which was less than the resolution of the OSA. In addition, the maximum output power fluctuation of the EDFL was 0.01 dB, which was equal to the resolution of the optical power meter. The above measured results indicate that the proposed SLM EDFL achieves excellent stability performance, which benefits from the all PM structure of the EDFL and the outstanding SLM selecting capability of the DR-PSR.

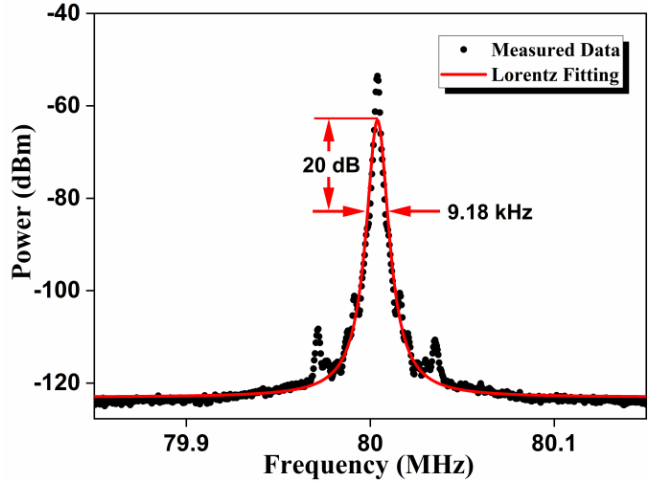

Fig. 7 The measured output spectrum of the laser linewidth.

To study the linewidth of the EDFL, the laser output linewidth was measured by the delayed self-heterodyne system [24] mentioned above. As shown in Fig. 7, the linewidth signal at 80 MHz was captured in a frequency range of 79.85 MHz~80.15 MHz with a 200 Hz RBW. By fitting the measured data with Lorentzian function, the 9.18 kHz of 20-dB bandwidth was obtained, as shown in the dotted line in Fig. 7. The measured data fitted well with the Lorentzian function and the linewidth of the EDFL could be estimated to be 459 Hz from the 20-dB spectrum width of the corresponding fitting curve. It should be noticed that the resolution of the delayed self-heterodyne system is limited by the 47 km SMF delay line since the fitting curve has the unavoidable broadening effect. In fact, we believe that the actual linewidth of the SLM EDFL should be narrower than 459 Hz.

Finally, the polarization characteristic of the all-PM EDFL was investigated with a polarization analyzer (PSY-201). As shown in Fig. 8(a), when the polarization characteristic of the EDFL was measured without vibration, the proposed EDFL output only one polarization state with an average degree of polarization (DOP) of 100.87%. In fact, the DOP should not be more than 100%, and the mirror analysis error is induced by the polarization analyzer. In order to investigate whether the all-PM SLM EDFL has the resistance capability to external disturbances, we gently flapped the fiber and the optical devices of the proposed EDFL, and the state of polarization measurement result is shown in Fig. 8(b). As shown in Fig. 8(b), there was almost only one polarization state of the EDFL with an average DOP of 100.84%. The state of polarization measurement results of the EDFL indicate that the proposed EDFL is insusceptible to the environmental perturbation, which mainly benefits from the all-PM structure of the EDFL.

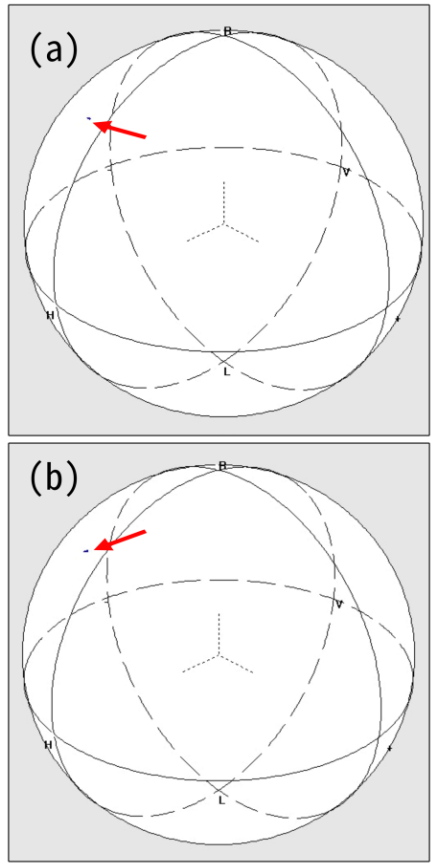

Fig. 8 States of polarization measurement results of the EDFL (a) without disturbance in 100-ms interval over 10 min and (b) with external disturbance in 100-ms interval over 1 min.

## 4. Conclusion

We have proposed and experimentally demonstrated an all-PM SLM EDFL with an ultra-high OSNR, an ultra-narrow linewidth and extremely high stability. A DR-PSR, composed of two single-coupler fiber rings and a length of unpumped EDF, is utilized as a high-quality filter to broaden the effective FSR of the compound cavity and to guarantee the SLM operation simultaneously. With the employment of the DR-PSR and the all-PM structure, we developed an all-PM SLM EDFL with excellent output performance and with strong ability to resist the environment disturbance. In the SLM operating mode, the OSNR value is as high as ~83 dB, and the laser linewidth is as narrow as 459 Hz at the pump power of 100 mW. Moreover, at the same pump level, the proposed EDFL possesses the excellent stability of both the lasing wavelength and the output power.

*Funding*

This paper is supported by Natural Science Foundation of China (NSFC) (61690195, 61701040 and 61531003); Beijing University of Posts and Telecommunications (BUPT) Action Project for Promoting the Development of Scientific and Technological Innovation (2019XD-A18); BUPT Excellent Ph.D. Students Foundation (XTCX201836 and CX2019107).

*Disclosures*

The authors declare no conflicts of interest.